\documentclass[iop]{emulateapj}
\usepackage{mathrsfs}
\usepackage{url}
\usepackage[normalem]{ulem}

\usepackage{natbib}
\usepackage{color}
\bibpunct{(}{)}{;}{a}{}{,}

\makeatletter

\newenvironment{figurehere}
  {\def\@captype{figure}}
  {}
\makeatother

\newcommand\aproxgt{\mathrel{%
      \rlap{\raise 0.511ex \hbox{$>$}}{\lower 0.511ex \hbox{$\sim$}}}}
\newcommand\aproxlt{\mathrel{%
      \rlap{\raise 0.511ex \hbox{$<$}}{\lower 0.511ex \hbox{$\sim$}}}}
\def\ir1334{{IRAS\,13349+2438}}
\def\mcg6{{MCG--6-30-15}}

\def\chandra{{\it Chandra}}

\def\lya{Ly$\alpha$}
\def\lyb{Ly$\beta$}

\def\kmps{\ifmmode \rm km~s^{-1} \else $\rm km~s^{-1}$\fi}
\def\psqcm{\ifmmode \rm cm^{-2} \else $\rm cm^{-2}$\fi}
\def\Msun{\ifmmode \rm M_{\odot} \else $\rm M_{\odot}$\fi}
\def\Lsun{\ifmmode \rm L_{\odot} \else $\rm L_{\odot}$\fi}

\newcommand{\qo}{\ifmmode q_{\rm o} \else $q_{\rm o}$\fi}
\newcommand{\Ho}{\ifmmode H_{\rm o} \else $H_{\rm o}$\fi}
\newcommand{\ho}{\ifmmode h_{\rm o} \else $h_{\rm o}$\fi}

\def\fake2{\hphantom{3}}

\shorttitle{A UV Counterpart to an Ultra-Fast X-ray Outflow in PG1211+143}
\shortauthors{Kriss et al.}

\begin{document}

\slugcomment{Accepted for publication in ApJ, 12/23/2017}

\title{Discovery of an Ultraviolet Counterpart to an Ultra-Fast X-ray Outflow in the Quasar PG1211+143}

\author{
Gerard~A.~Kriss\altaffilmark{1},
Julia~C.~Lee\altaffilmark{2,3},
Ashkbiz Danehkar\altaffilmark{2},
Michael A. Nowak\altaffilmark{4},
Taotao~Fang\altaffilmark{5},
Martin~J.~Hardcastle\altaffilmark{6},
Joseph~Neilsen\altaffilmark{7},
Andrew~Young\altaffilmark{8}
}

\altaffiltext{1}{Space Telescope Science Institute, 3700 San Martin
Drive, Baltimore, MD 21218, USA}
\altaffiltext{2}{Harvard-Smithsonian Center for Astrophysics, 60 Garden Street, Cambridge, MA 02138, USA}
\altaffiltext{3}{Harvard University, John A. Paulson School of Engineering \& Applied Science, 29 Oxford Street, Cambridge, MA 02138, USA}
\altaffiltext{4}{Massachusetts Institute of Technology, Kavli
Institute for Astrophysics, Cambridge, MA 02139, USA}
\altaffiltext{5}{Xiamen University, Department of Astronomy, Xiamen, Fujian 361005, P. R. CHINA}
\altaffiltext{6}{University of Hertfordshire, School of Physics, Astronomy and Mathematics, Hatfield, Hertfordshire AL10 9AB, UK}
\altaffiltext{7}{Villanova University, Mendel Hall, Room 263A,
800 E. Lancaster Avenue, Villanova, PA 19085, USA}
\altaffiltext{8}{University of Bristol, School of Physics, HH Wills Physics Laboratory, Bristol BS8 1TH, UK}

\begin{abstract} 
We observed the quasar PG1211+143 using the Cosmic Origins Spectrograph on the
Hubble Space Telescope in April 2015 as part of a joint campaign with the
{\it Chandra} X-ray Observatory and the Jansky Very Large Array.
Our ultraviolet spectra cover the wavelength range 912--2100 \AA.
We find a broad absorption feature ($\sim 1080~\kmps$) at an observed wavelength
of 1240 \AA. Interpreting this as \ion{H}{1} Ly$\alpha$,
in the rest frame of PG1211+143 ($z=0.0809$), this corresponds to an
outflow velocity of $-16\,980~\rm km~s^{-1}$
(outflow redshift $z_{\rm out} \sim -0.0551$),
matching the moderate ionization X-ray absorption system detected in our
{\it Chandra} observation and reported previously by Pounds et al. (2016).
With a minimum H\,{\sc i} column density of log $\rm N_{HI} > 14.5$,
and no absorption in other UV resonance lines, this Ly$\alpha$ absorber is
consistent with arising in the same ultra-fast outflow as the X-ray
absorbing gas.
The Ly$\alpha$ feature is weak or absent in
archival ultraviolet spectra of PG1211+143, strongly suggesting that this
absorption is transient, and intrinsic to PG1211+143.
Such a simultaneous detection in two independent wavebands
for the first time
gives strong confirmation of the reality of an ultra-fast outflow
in an active galactic nucleus.
\end{abstract}


%%%%%%%%%%%%%%%%%%%%%%%%%%%%%%%%%%%%%%%%%%%%%%%%%%%%%%%%%%%%%%%%%%%%%%%%%%%%%
%%%%%%%%%%%%%%%%%%%%%%%%%%%%%%%%%%%%%%%%%%%%%%%%%%%%%%%%%%%%%%%%%%%%%%%%%%%%%
%%%%%%%%%%%%%%%%%% Main
%%%%%%%%%%%%%%%%%% Paper %%%%%%%%%%%%%%%%%%%%%%%%%%%%%%%%%%%%%%%%%%%%%%%

\section{Introduction}
\label{section:intro}
Fast, massive outflows from active galactic nuclei (AGN) may play a prominent
role in the evolution of their host galaxies. 
These outflows may both heat and remove the interstellar medium (ISM) of the
host galaxy, effectively stopping further star formation, and removing the fuel
for further black hole growth
\citep{Silk98,King03,Ostriker10,Soker10,Faucher12,Zubovas14,Thompson15}.
If the kinetic luminosity is high enough, 0.5\% \citep{Hopkins10} to 5\%
\citep{DiMatteo05} of the bolometric luminosity, then the impact on the
host galaxy may be sufficient to regulate galaxy growth and produce the
observed $\rm M_{BH} - \sigma_{\rm bulge}$ correlation
\citep{Ferrarese00, Gebhardt00}.
Recent observations of high-luminosity AGN at moderate redshifts demonstrate
that outflows of this power spanning galactic scales
do exist \citep{Borguet13}, and that such outflows may be ubiquitous, even
when not seen in absorption along the line of sight
\citep{Liu13a, Liu13b, Liu14}.

Outflows implied by the X-ray warm absorbers and blue-shifted UV absorption
lines commonly seen in nearby AGN \citep{Crenshaw03} are often too weak to
potentially influence their host galaxies \citep{Crenshaw12}.
On the other hand, ultra-fast outflows (UFOs),
typified by high column densities of highly ionized gas and primarily
identified via \ion{Fe}{26} K$\alpha$ absorption
outflowing at velocities of $> 10\,000~\kmps$ would have the mass and kinetic
energy to make a substantial impact on the evolution of their hosts
\citep{Pounds03,Tombesi10,Tombesi14b}.
However, given the low statistical significance of these features and the fact
that they are often based on the identification of only a single spectral
feature, \cite{Vaughan08} have questioned their reality.
The large, comprehensive Warm Absorbers in X-rays (WAX) survey \citep{Laha14}
found no significant statistical evidence for UFOs
in the six sources they had in common with \cite{Tombesi10}.
While not discussing the data per se, \cite{Gallo13} argue for an alternative
explanation based on blurred reflection rather than an outflowing wind.

\begin{figure*}
\centering
\includegraphics[width=0.99\textwidth]{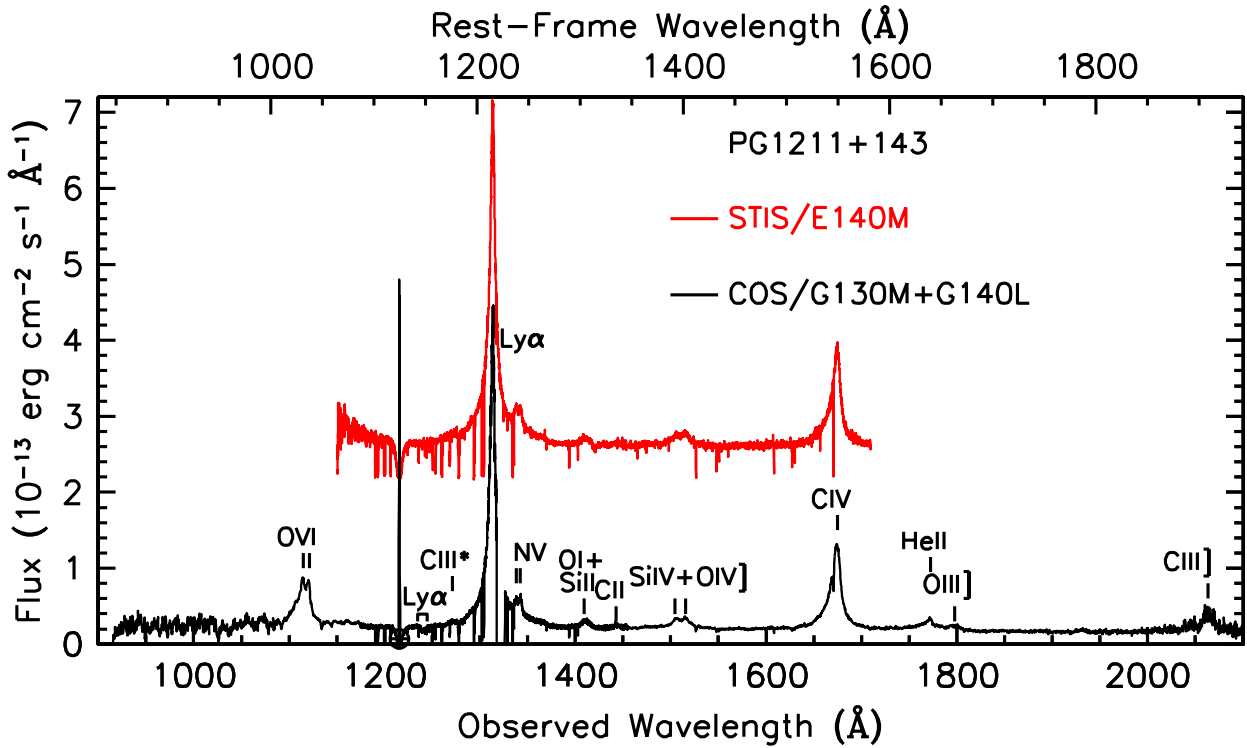}
\caption{Merged {\it HST}/COS G130M and G140L spectra of PG1211+143
(black)
compared to the 2002 STIS/E140M spectrum (red).
Wavelengths on the lower horizontal axis and fluxes are observed units.
The upper horizontal axis shows wavelengths in the rest frame of PG1211+143
at $z=0.0809$.
For clarity, the STIS spectrum has been offset vertically by
$2.2 \times 10^{-13}~\rm erg~cm^{-2}~s^{-1}~\AA^{-1}$.
Prominent emission lines are marked.
An earth symbol denotes the strong geocoronal Ly$\alpha$ emission line.
All narrow absorption features are either foreground ISM or IGM absorption lines.
The broad Ly$\alpha$ absorption feature
in the COS spectrum} is marked above the spectrum in the blue
wing of the Ly$\alpha$ emission line.
\label{FigFullSpectrum}
\end{figure*}

The Quasi-Stellar Object (QSO) PG1211+143 ($z = 0.0809$)
plays a central role in the controversy over
relativistic outflows because it presents tantalizing evidence
for the presence of UFOs and both intermediate and lower velocity flows
typical of warm absorbers.
\cite{Pounds03, Pounds16} identified two UFOs: one at high velocity
$v_{\rm out}\sim-0.14c,$ but also a lower velocity UFO at
$v_{\rm out}\sim-0.06c$\footnote{
We use the convention that the velocities given in prior work on PG1211+143 are
relativistic velocities in its rest frame, with $v_{\rm out}$ represented as
$zc$ simply by dividing $v_{\rm out}$ by $c=2.9979 \times 10^5~\rm km~s^{-1}$.}
However, studying the same original XMM observation of PG1211+143 as
\cite{Pounds03}, \cite{Kaspi06}
find no evidence for UFOs, but rather lower-velocity systems more typical of
those seen in warm absorbers.
\cite{Pounds09} and \cite{Tombesi11} find UFOs persistently present, but
varying in strength over the course of several XMM observations spanning months
to years, while a long, 300 ks NuSTAR observation of PG1211+143 also finds no
evidence for UFOs \citep{Zoghbi15}.

Prior UV spectra of PG1211+143 revealed the usual blue continuum and broad
emission lines typical of a Type 1 AGN, but no absorption lines typical
of the outflows seen in other AGN.
All absorption lines in the spectrum (including ones at velocities near that
of the $v_{\rm out} \sim -0.06~c$ X-ray absorber) were identified as
intervening gas in the intergalactic medium (IGM)
\citep{Penton04, Tumlinson05, Danforth08, Tilton12}.
Given the variety of results obtained on PG1211+143 and its prominence in the
controversy over the reality of high-velocity outflows in AGN, we undertook a
large joint campaign using the {\it Chandra} X-ray Observatory,
the {\it Hubble Space Telescope} ({\it HST}), and
the Karl G. Jansky Very Large Array (VLA)
to search for both X-ray and ultraviolet outflowing absorption systems.
We report on the {\it HST} UV results here.
See \cite{Danehkar17} for the {\it Chandra} HETGS X-ray results
complementing this paper.
As in \cite{Danehkar17}, in this paper we will use the following conventions for
velocities and redshift:

$z_{\rm rest}=0.0809$ defines the rest frame of PG1211+143.

$z_{\rm obs}$ is the observed redshift of a spectral feature
in our reference frame.

$z_{\rm out}$ gives the redshift of an outflow in the rest frame of PG1211+143.

$v_{\rm out}$ gives the velocity of an outflow in the rest frame of PG1211+143.

$\lambda_{\rm obs}$ is the observed wavelength of a spectral feature.

$\lambda_0$ is the rest wavelength (vacuum) of a spectral feature.

The usual special relativistic relations are used for conversions among
the various quantities:

$z_{\rm obs} = (\lambda_{\rm obs}/\lambda_0) - 1$,

$z_{\rm out} = (1+z_{\rm obs})/(1+z_{\rm rest}) - 1$,

$v_{\rm out} = c [(1+z_{\rm out})^2-1]/[(1+z_{\rm out})^2+1]$, and

$z_{\rm out} = \sqrt{[(1+v_{\rm out}/c)/(1-v_{\rm out}/c)]} - 1$,
 where $c$ is the speed of light.

\section{{\it HST} Observations}

In April 2015 we observed PG1211+143 using {\it Chandra} and {\it HST} in a
coordinated set of visits. The {\it HST}/COS observations used grating G140L with a
central wavelength setting of 1280 to cover the entire 912--2000 \AA\ wavelength
range \citep{Green12}.
To fill in the gap in wavelength coverage between segments A and B of the
COS detector, in our second visit we also used grating G130M with a central
wavelength setting of 1327 \AA. The observations are summarized in
Table \ref{HSTObsTbl}. All observations were split into four exposures at
different FP-POS positions to enable us to remove detector artifacts and
flat-field features.

\begin{table*}
  \caption[]{{\it HST}/COS Observations of PG1211+143}
\label{HSTObsTbl}
\begin{center}
\begin{tabular}{l c c c c c}
\hline\hline
Proposal & Data Set & Grating/Tilt  & Date & Start Time & Exposure\\
   ID         & Name &           &         &    (GMT)   & (s)\\
\hline
13947 & lcs501010 & G140L/1280 & 2015-04-12 & 15:50:03 & 1900 \\
13947 & lcs504010 & G140L/1280 & 2015-04-14 & 13:52:21 & 1900 \\
13947 & lcs502010 & G140L/1280 & 2015-04-14 & 15:36:39 & 1900 \\
13947 & lcs502020 & G130M/1327 & 2015-04-14 & 17:16:34 & 2320 \\
\hline
\end{tabular}
\end{center}
\end{table*}

\begin{figurehere}
\centering
\includegraphics[angle=-90, width=0.52\textwidth]{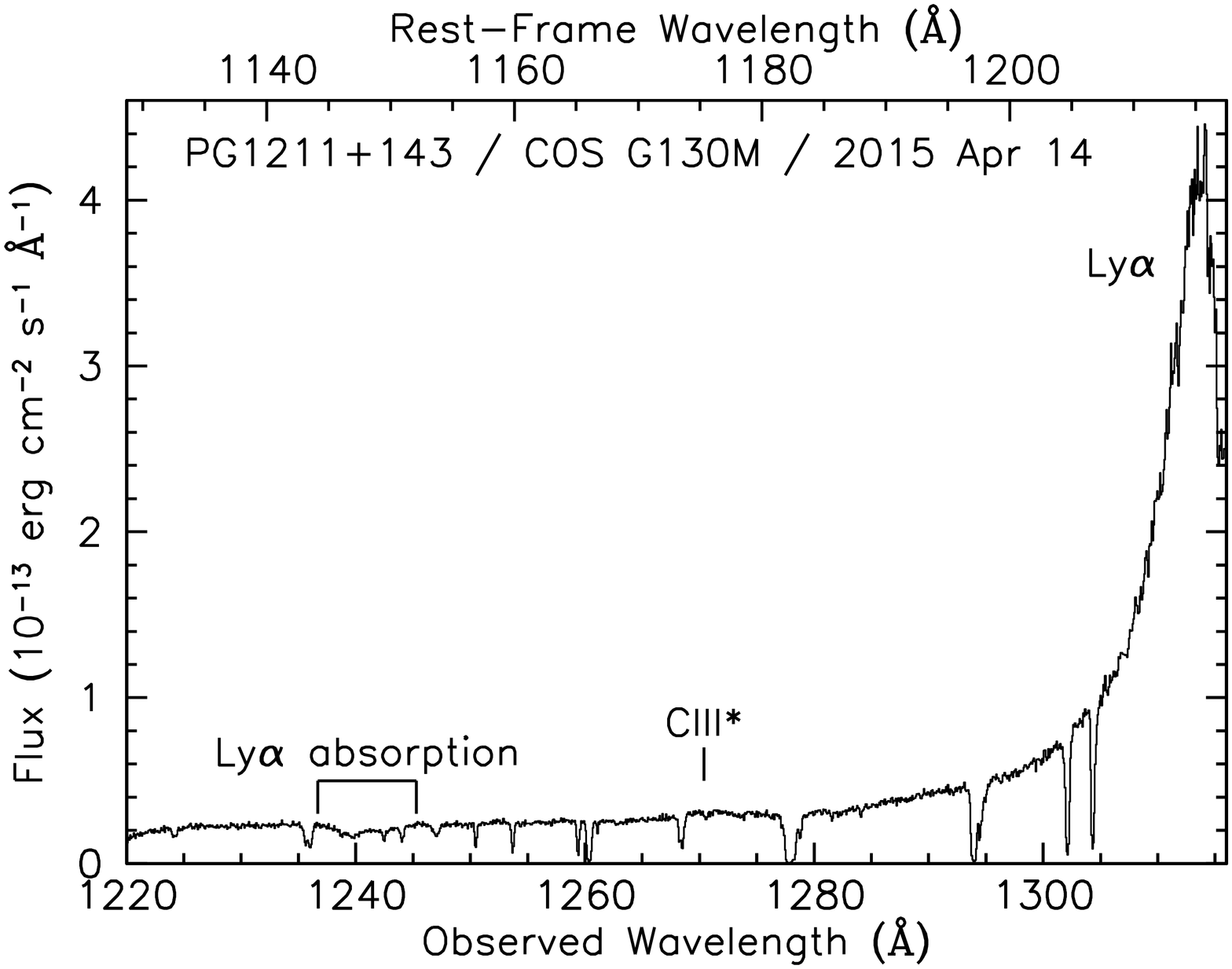}
\caption{{\it HST}/COS G130M spectrum covering wavelengths in the
blue wing of the Ly$\alpha$ emission line of PG1211+143.
The lower horizontal axis is the observed wavelength in \AA ngstroms.
The upper horizontal axis gives wavelengths in the rest frame of PG1211+143
at $z = 0.0809$.
The broad feature labeled ``Ly$\alpha$ absorption" at an observed wavelength
of 1240 \AA\ is weak or absent in archival spectra.
Emission lines of \ion{C}{3}* $\lambda 1176$ and Ly$\alpha$ in PG1211+143 are
labeled. All other narrow absorption lines arise in foreground interstellar
or intergalactic gas.
}
\label{FigG130M}
\end{figurehere}
\vskip 12pt

The individual exposures in our program were combined by grating with updated
wavelength calibrations, flat-fields, and flux calibrations using the methods
of \cite{Kriss11b} and \cite{DeRosa15}.
To adjust the wavelength zero points of our spectra, for G130M, we
cross-correlated our spectra with the archival STIS spectrum of PG1211+143
\citep{Tumlinson05}. For G140L, we measured the wavelengths of low-ionization
interstellar lines and molecular hydrogen features and compared them to the
\ion{H}{1} velocity of $v_{\rm LSR} = -17~\rm km~s^{-1}$ \citep{Wakker11}.
No adjustment to the G140L wavelengths was required.

Our {\it HST} observations showed PG1211+143 to be similar in appearance to archival
{\it HST} and IUE observations as shown in Figure \ref{FigFullSpectrum}.
The continuum flux at 1350 \AA\ rest (1465 \AA\ observed) was
$f_\lambda = 2.2 \times 10^{-14}~\rm erg~cm^{-2}~s^{-1}~\AA^{-1}$,
slightly below the historical median flux of
$2.9 \times 10^{-14}~\rm erg~cm^{-2}~s^{-1}~\AA^{-1}$.
Despite the lower flux and our shorter observation time, the signal-to-noise
ratio (S/N) of our observation ($\sim 29$ per resolution element for G130M at
1240 \AA) significantly improved upon the prior 25-orbit STIS echelle spectrum.
Since the goal of our observations was to look for evidence of outflowing gas
in PG1211+143 as evidenced by blue-shifted absorption lines, we scrutinized
our spectra carefully.
This revealed a previously unknown weak, broad feature in the blue wing of the
\lya\ emission line as shown in Figure \ref{FigG130M} and
the upper panel of \ref{FigLya}.

\begin{figurehere}
\centering
\includegraphics[angle=0, width=0.52\textwidth]{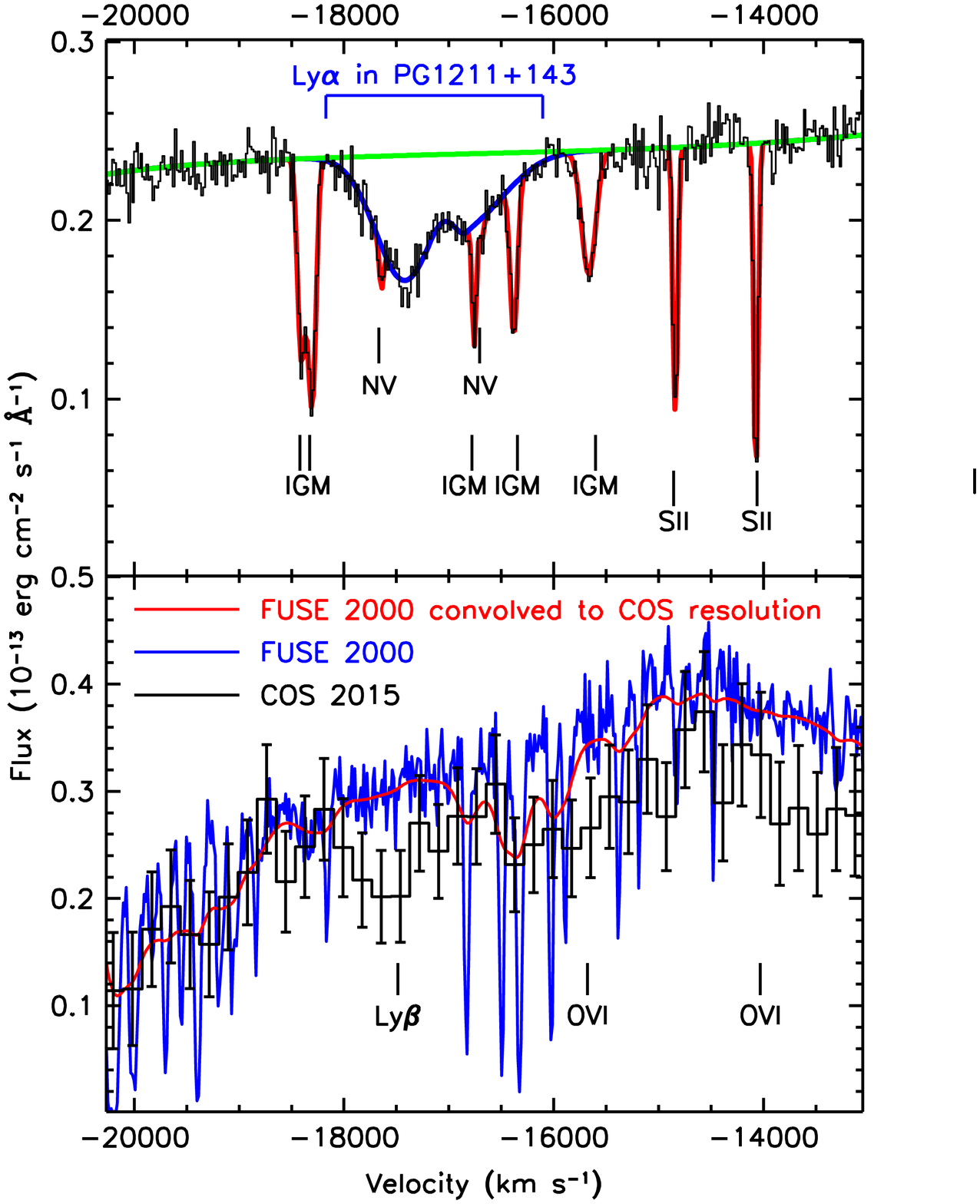}
\caption{
Upper panel: {\it HST}/COS G130M spectrum of PG1211+143 in the wavelength region
surrounding the broad Ly$\alpha$ absorption feature.
The horizontal axis gives the outflow velocity relative to Ly$\alpha$
in the rest frame of PG1211+143 at $z = 0.0809$.
The blue line is our best-fit model for the broad Ly $\alpha$ absorption
intrinsic to PG1211+143.
Intergalactic Ly$\alpha$ lines
identified by \cite{Penton04} are indicated by tick marks labeled ``IGM".
Interstellar lines of \ion{N}{5} and \ion{S}{2} are also labeled.
The red line shows our best-fit model for these foreground absorption lines.
The green line shows the emission model (continuum plus broad Ly $\alpha$
emission) with all absorption removed.
Lower Panel:
COS and FUSE spectra of PG1211+143 in the wavelength region
corresponding to broad Ly$\beta$ absorption.
The horizontal axis gives the outflow velocity relative to Ly$\beta$
in the rest frame of PG1211+143 at $z = 0.0809$.
The black histogram with 1-$\sigma$ error bars shows the COS G140L data,
binned by 8 pixels.
The blue line is the FUSE spectrum from 2000 \citep{Tumlinson05}.
All the absorption features in the FUSE spectrum are foreground interstellar
absorption.
The red line is the FUSE data convolved with the COS G140L line spread function.
Both the original FUSE data and the convolved spectrum are scaled to the
intensity of the COS spectrum at 1040 \AA.
The expected minimum of Ly$\beta$ absorption that would correspond to the
Ly$\alpha$ absorption trough is marked.
The expected locations of \ion{O}{6} $\lambda\lambda 1032,1038$
are also marked.
}
\label{FigLya}
\end{figurehere}
\vskip 12pt

Although weak, narrow
interstellar and intergalactic absorption lines have been previously cataloged
in this region \citep{Penton04, Tumlinson05, Danforth08, Tilton12},
this broad dip centered at $\sim 1240$ \AA\ was
not readily visible in prior {\it HST} spectra.
To convince ourselves that this feature was intrinsic to PG1211+143, and not an
artifact in COS, we examined the individual exposures in each FP-POS setting.
The broad absorption feature appears in all four exposures.
Furthermore, no similar feature is present in any of the white dwarf standard
star spectra obtained monthly as part of the COS calibration monitoring
program.

To conclusively associate this single spectral feature with \lya\ absorption
intrinsic to PG1211+143, we note:
(1) If it were \ion{N}{5}, one would expect to see \lya\ at shorter
wavelengths, and also \ion{C}{4} at longer wavelengths.
No features are present at those expected wavelengths in our observations.
(2) The velocity of this feature in the rest frame of PG1211+143 matches the
velocity of the detected soft X-ray absorption \citep{Danehkar17}.
(3) We also detect \lyb\ as described later in this section.

To measure the strength of the \lya\ absorption feature,
we used {\tt specfit} \citep{Kriss94} in IRAF to model the surrounding continuum
and line emission and the embedded ISM and IGM absorption lines.
For the continuum we used a reddened power law of the form
$\rm f_\lambda = 3.78 \times 10^{-14} (\lambda/1000 \AA)^{-0.779}
~\rm erg~cm^{-2}~s^{-1}~\AA^{-1}$
with foreground Galactic extinction of
$\rm E(B - V)$ = 0.030 \citep{Schlafly11}.
We also included foreground
damped \lya\ absorption due to the ISM with a column density of
$\rm N_H = 2.588 \times 10^{20}~cm^{-2}$ \citep{Wakker11}.
The \lya\ emission line of PG1211+143 was modeled using three Gaussian emission
components. Their parameters are summarized in Table \ref{LyaEmTbl}.
% Since the broad absorption we detect is in the far blue wing of \lya, the
% details of the \lya\ emission model are inconsequential.
For the narrow foreground ISM and IGM lines, we used individual Voigt profiles.
Finally, for the broad intrinsic \lya\ absorption, we modeled its profile using
two blended Gaussians in negative flux in order to account for its asymmetric,
uneven profile.
Since the broad  \lya\ line is well resolved, we obtain a lower limit on the
column density using the apparent optical depth method \citep{Savage91}
by integrating over the normalized absorption profile.
The measured properties of the broad \lya\ absorber are summarized in
Table \ref{AbsTbl}. Here we give the properties of the individual components
in our fit as well as the properties of the full blended trough.

\begin{table*}
  \caption[]{Parameters of the Broad Emission Components in PG1211+143}
  \label{LyaEmTbl}
\begin{center}
\begin{tabular}{l c c c c c c}
\hline\hline
{Line} & $\lambda_o$ & {Flux} & {Velocity} & {FWHM} \\
      & (\AA)       & ($\rm 10^{-14}~erg~cm^{-2}~s^{-1}$) & ($\rm km~s^{-1}$) & ($\rm km~s^{-1}$) \\
\hline
{STIS 2013} \\
\hline
\ion{C}{3}*     & 1176.0   & $\phantom{00}3.5 \pm 0.3$   & $\phantom{0}-560 \pm 44$ & $\phantom{00}1400 \pm 130$ \\
Ly$\alpha$ & 1215.67  & $\phantom{00}9.6 \pm 0.3$   & $\phantom{0}1240 \pm 40$ & $\phantom{00}460 \pm 170$ \\
Ly$\alpha$ & 1215.67  & $\phantom{00}33 \pm 0.2$   & $\phantom{0}-170 \pm 5$ & $\phantom{00}660 \pm 10$ \\
Ly$\alpha$ & 1215.67  & $280 \pm 0.4$   & $\phantom{0-}300 \pm 5$ & $\phantom{0}2200 \pm 10$ \\
Ly$\alpha$ & 1215.67  & $110 \pm 0.2$   & $\phantom{}-1070 \pm 5$ & $\phantom{0}3800 \pm 30$ \\
Ly$\alpha$ & 1215.67  & $250 \pm 0.1$   & $\phantom{0-}980 \pm 5$ & $13800 \pm 22$ \\
\hline
{COS 2015} \\
\hline
\ion{C}{3}* & 1176.0   & $\phantom{00}0.8 \pm 0.3$   & $\phantom{0}-160 \pm 190$ & $\phantom{00}860 \pm 120$ \\
Ly$\alpha$ & 1215.67  & $\phantom{00}7.4 \pm 0.9$   & $\phantom{0}-180 \pm 17$ & $\phantom{00}330 \pm 26$ \\
Ly$\alpha$ & 1215.67  & $\phantom{00}7.6 \pm 1.2$   & $\phantom{00-}20 \pm 160$ & $\phantom{0}1000 \pm 36$ \\
Ly$\alpha$ & 1215.67  & $143 \pm 1.8$   & $\phantom{0}-100 \pm 5$ & $\phantom{0}1450 \pm 14$ \\
Ly$\alpha$ & 1215.67  & $142 \pm 0.5$   & $\phantom{0}-470 \pm 6$ & $\phantom{0}3600 \pm 26$ \\
Ly$\alpha$ & 1215.67  & $196 \pm 1.5$   & $\phantom{-}860 \pm 30$ & $13100 \pm 33$ \\
\hline
\end{tabular}
\end{center}
\end{table*}

\begin{table*}
  \caption[]{Properties of the Broad Ly$\alpha$ Absorption in PG1211+143}
  \label{AbsTbl}
\begin{center}
\begin{tabular}{l c c c c c c c}
\hline\hline
{Line} & $\lambda_o$ & {EW} & {Velocity} & {FWHM} & $\rm C_f$ & log $\rm N_{ion}$ & {Predicted log $\rm N_{ion}$}\\
      & (\AA)       & (\AA) & ($\rm km~s^{-1}$) & ($\rm km~s^{-1}$) & & (log $\rm cm^{-2}$) &  ($\rm log~cm^{-2}$) \\
\hline
{FOS 1991} \\
\hline
Ly$\alpha$ & 1215.67  & $<0.45$ & $-16\,980$ & $1080$ & 1.0 & $<13.90$ & $\ldots$ \\
\hline
{STIS 2002} \\
\hline
Ly$\alpha$ & 1215.67  & $0.45 \pm 0.04$ & $-17\,450 \pm 15$ & $880 \pm 10$ & 0.1 & $<13.90$ & $\ldots$ \\
\hline
{COS 2015} \\
\hline
Ly$\alpha$ & 1215.67  & $0.86 \pm 0.12$ & $-17\,420 \pm 15$ & $653 \pm 36$ & $0.30 \pm 0.04$ & $>14.30$ & $\ldots$ \\
Ly$\alpha$ & 1215.67  & $0.41 \pm 0.17$ & $-16\,825 \pm 36$ & $320 \pm 74$ & $0.30 \pm 0.04$ & $>13.95$ & $\ldots$ \\
Ly$\alpha$ total & 1215.67  & $1.27 \pm 0.18$ & $-16\,980 \pm 40$ & $1080 \pm 800$ & $0.30 \pm 0.04$ & $>14.46$ & 13.95 \\
Ly$\beta$ & 1025.72  & $0.91 \pm 0.27$ & $-17\,464 \pm 90$ & $350 \pm 50$ & $0.33 \pm 0.14$ & $>15.20$ & 13.95 \\
\hline
\end{tabular}
\end{center}
\end{table*}

Since the short wavelength segment of our G140L grating observations covers
the \lyb\ region of PG1211+143, we are also able to measure \lyb\ absorption
over the same velocity range as we see in \lya.
The spectrum in this observed wavelength range is more complex due to
foreground Galactic ISM features and the lower resolution of the G140L grating.
To aid in this analysis, we retrieved the archival FUSE observation of
PG1211+143 reported by \cite{Tumlinson05}. We convolved this with the
COS G140L line-spread function \citep{Duval13}
and scaled the flux level to match
our COS spectrum at 1060 \AA. The comparison shown in
the lower panel of Figure \ref{FigLya}
reveals a deficiency in flux in our COS spectrum relative to FUSE
precisely at the wavelengths expected for a \lyb\ counterpart to the G130M
\lya\ absorption feature.
We are not able to resolve the \lyb\ absorption in the same detail as we can
for \lya, so we simply measure its integrated properties.
If we use the scaled and convolved FUSE spectrum
to normalize the COS spectrum and integrate this normalized spectrum over the
velocity range $-1500$ to $+1500~\rm km~s^{-1}$,
again using the apparent optical depth method of \cite{Savage91},
we obtain an equivalent width
(EW) of $0.91 \pm 0.27$ \AA.
This flux deficiency is significant at a confidence level of $> 0.998$
compared to the null hypothesis of no absorption.
As shown in Table \ref{AbsTbl}, the strength of the \lyb\ absorption is
comparable to that of \lya, suggesting that both spectral features might
be heavily saturated.
This sets a lower limit on the \ion{H}{1} column density of
$\rm N_H > 2.9 \times 10^{14}~cm^{-2}$.
Given that the features are not black at the bottoms of the troughs,
the \ion{H}{1} absorption would then only partially cover the continuum
source. To measure the the covering fractions cited in Table \ref{AbsTbl},
we use the depth at the center of the absorption trough.
We note that since the \lyb\ feature appears to be narrower than \lya, this
may indicate that the absorber is stratified in its column density, being more
optically thick at line center than at higher velocities.

We do not see absorption associated with other high-ionization lines in the
COS spectrum. No troughs associated with \ion{O}{6}, \ion{N}{5}, or \ion{C}{4}
are visible in our spectra.
Although there appear to be deficiencies in flux in the COS spectrum at the
expected locations of the \ion{O}{6} doublet in the lower panel of
Figure \ref{FigLya}, these are not statistically
significant at more than 2$\sigma$ confidence.
Table \ref{tab:limits} gives upper limits at 2$\sigma$ confidence
to the equivalent widths and column densities of these features
assuming they have profiles similar to the detected Ly$\alpha$ absorption.
The high saturation present in \ion{H}{1} and this
lack of associated high-ionization lines suggests that this absorbing gas is
both very highly ionized and of high total column density.

\begin{table*}
  \caption[]{Upper Limits for Absorption Features in PG1211+143}
  \label{tab:limits}
\begin{center}
\begin{tabular}{l c c c c c c c}
\hline\hline
{Line} & $\lambda_o$ & {EW} & {Velocity} & {FWHM} &  $\rm log~N_{ion}$ & Predicted $\rm log~N_{ion}$\\
      & (\AA)       & (\AA) & ($\rm km~s^{-1}$) & ($\rm km~s^{-1}$) & ($\rm log~cm^{-2}$) &  ($\rm log~cm^{-2}$) \\
\hline
\ion{O}{6} & 1032,1038  & $<0.62\phantom{0}$ & -16\,980 & 1080 & $<14.51$ & $12.83$ \\
\ion{N}{5} & 1238,1242  & $<0.12\phantom{0}$ & -16\,980 & 1080 & $<13.58$ & $10.72$ \\
\ion{C}{4} & 1548,1550  & $<0.22\phantom{0}$ & -16\,980 & 1080 & $<13.56$ & $\phantom{0}8.91$ \\
Ly$\alpha$ & 1215.67    & $<0.17\phantom{0}$ & \phantom{0}-3\,000 & 1080 & $<13.50$ & $\ldots$ \\
Ly$\alpha$ & 1215.67    & $<0.074$ & -24\,000 & 1080 & $<13.13$ & $\ldots$ \\
Ly$\alpha$ & 1215.67    & $<0.53\phantom{0}$ & -38\,700 & 1080 & $<14.00$ & $14.49$ \\
\hline
\end{tabular}
\end{center}
\end{table*}

\cite{Tumlinson05} observed PG1211+143 using a very deep (45 ks) STIS echelle
E140M observation in 2002. We examined this archival spectrum to see if there
was any indication of \lya\ absorption  at the velocity of our COS detection.
Figure \ref{FigLyaVary} compares the prior STIS observation of PG1211+143 to
our new COS observation. One can see a slight depression in the same region as
the much more prominent absorption we have detected with COS. Although this
depression is marginally significant (P $> 0.96$ for the null hypothesis of
no absorption), it is not an obvious feature one would have selected without
knowing where to look in the spectrum. Its weakness (or even absence) in the
prior STIS spectrum indicates that this \ion{H}{1} absorption feature is
variable in strength.
The {\it HST}-Faint Object Spectrograph observation of PG1211+143 in 1991 also
bolsters this case for variability. Here we can set an upper limit on the
presence of a Ly$\alpha$ absorption feature at $v_{out} = -16\,980~\rm km~^{-1}$
comparable to the strength of that in the STIS spectrum.

Outflows at velocities of $-3000~\rm km~s^{-1}$ and  $-24,000~\rm km~s^{-1}$
have also been reported in prior X-ray observations of PG1211+143
\citep{Pounds03, Kaspi06}.
We have carefully examined our COS spectra in these velocity ranges.
As shown in Table \ref{tab:limits}, we find
no evidence for \ion{H}{1} absorption at any velocity other than surrounding
$-16\,980~\rm km~s^{-1}$.

\section{Discussion}

Our joint {\it Chandra} and {\it HST} observations of PG1211+143 clarify
the confusing kinematics of at least one major outflow
component in this important example of a UFO.
The {\it HST}-COS detection of a broad Ly$\alpha$ absorption feature at an
outflow velocity of $-$16\,980 $\rm km~s^{-1}$ ($0.0551 c$)
matches the velocity of the
high-ionization absorption component detected in the joint {\it Chandra}-HETGS
spectrum at $v_{\rm out} = -17\,300~\rm km~s^{-1}$
($z_{\rm out} = -0.0561 c$) \citep{Danehkar17}.
This absorber may be the same as the $-0.066 c$ component detected in the
deeper {\it XMM-Newton} EPIC-pn observation of \cite{Pounds16}, but it has
lower ionization, log $\xi = 2.81$ compared to log $\xi = 3.4$,
and it is much lower in column density,
$\rm N_H = 3 \times 10^{21}~cm^{-2}$
compared to $\rm 2 \times  10^{23}~cm^{-2}$.
Analysis of the RGS data from the 2014 XMM-{\it Newton} observations
\citep{Reeves17} reveals that this
absorber has two components at velocities of
$-0.062 \pm 0.001 c$ ($v_{out} = -18\,600 \pm 300~\rm \kmps$) and
$-0.059 \pm 0.002 c$ ($v_{out} = -17\,700 \pm 600~\rm \kmps$), the latter of
which is compatible with our detected \ion{H}{1} absorption.

The kinematics of the \chandra\ X-ray absorber detected by \cite{Danehkar17}
make it a good match to the
{\it HST}-COS Ly$\alpha$ absorber reported in this paper.
A crucial question, however, is whether the absorbing gas detected in our UV
spectrum is identically the same gas seen in the \chandra\ spectrum with the
exact same physical conditions.
\cite{Fukumura2010b} constructed a photoionization model of a
magnetohydrodynamically accelerated UFO in which the high-ionization gas
producing \ion{Fe}{25} could also have associated UV absorption lines 
(\ion{C}{4} in particular).
They find that producing detectable ionic concentrations of low-ionization
species typical of UV spectra, e.g., \ion{C}{4}, in such high-ionization gas
requires a fairly soft spectrum with a low X-ray to UV luminosity ratio.
In their model, they require $\alpha_{ox} = 1.7$, which is characteristic of
higher redshift, high-luminosity QSOs.
In the $z = 3.912$ UFO source APM~08279+5255, which has $\alpha_{ox} = 1.7$,
\cite{Hagino2017} successfully produce a model that includes both
low-ionization UV absorption consistent with the broad UV absorption lines
viewed in this object as well as lower-ionization X-ray absorption
(compared to \ion{Fe}{25}).
In contrast, our {\it Chandra}+{\it HST} observations show that
PG~1211+143 has a much higher X-ray to UV luminosity ratio, with an
observed $\alpha_{ox} = 1.47$. 
Our best-fit photoionization model for the X-ray absorbing gas
in \cite{Danehkar17} predicts very low column densities for all commonly
observed UV metal ions (\ion{C}{4}, \ion{N}{5}, and \ion{O}{6}).
These predicted column densities are given in the last column of
Table \ref{tab:limits}, and they are far below the upper limits for these ions
that we measure in our {\it HST} spectra.

Although the ionic concentrations of the UV metal ions are extremely low,
the predicted column density of \ion{H}{1} is much higher due to its
high abundance.
Our
best-fit photoionization model \citep{Danehkar17}
predicts a neutral hydrogen column of $8.8 \times 10^{13}~\rm cm^{-2}$.
This is lower than the lower limit derived from our \lya\ measurement,
$>2.9 \times 10^{14}~\rm cm^{-2}$, but
this prediction hinges crucially on the shape of the ionizing spectrum in the
Lyman continuum.
Our assumed spectral energy distribution
is weighted toward a
high ionizing luminosity since it extrapolates both the UV continuum and the
soft X-ray continuum to a meeting point in the extreme ultraviolet
\citep[Figure 4 in][]{Danehkar17}. However, a softer SED with a break to a
steeper power law at the Lyman limit that then extrapolates to the
detected soft X-ray continuum has half the ionizing flux in the Lyman
continuum. Spectra with such a break are common in composite quasar spectra
\citep{Zheng97, Telfer02} and in the spectra of individual objects
\citep{Shang05}. With such a softer SED, the predicted neutral hydrogen
column is $3.2 \times 10^{14}~\rm cm^{-2}$, which is compatible with our
UV observation.

Alternatively, one could reconcile the lower predicted \ion{H}{1} column
density of the X-ray spectrum with the higher column density observed in
\lya\ if the X-ray absorber is associated with only a portion of the
\lya\ trough. As illustrated by our model in Fig. \ref{FigLya} and the
parameters in Table \ref{AbsTbl},
the red component of the \lya\ blend has a lower column
density, compatible with the X-ray absorber. Its line width
(FWHM $= 320 \pm 74~\rm km~s^{-1}$, $v_{\rm turb} = 226 \pm 52~\rm km~s^{-1}$)
is also a better match to the turbulent velocity inferred for the X-ray
absorber, $v_{\rm turb} =  91^{+205}_{-59}~\rm km~s^{-1}$.

Although the column densities of detected (\ion{H}{1})
and undetected UV species are quite compatible with both
the X-ray and UV absorption arising in the same gas,
the complexity of the UV line profile relative to the X-ray may indicate
that there are physically separate zones commingled in the outflow.
\cite{Hagino2017} suggest that the UV absorbing gas in APM~08279+5255 is due
to higher-density clumps embedded in the X-ray UFO.
This may be true in PG~1211+143, but one would need
higher signal-to-noise ratio X-ray observations with better spectral resolution
as well as higher signal-to-noise ratio UV observations of the
Ly$\beta$ and \ion{O}{6} region to resolve this possibility.

\begin{figurehere}
\centering
\includegraphics[angle=-90, width=0.52\textwidth]{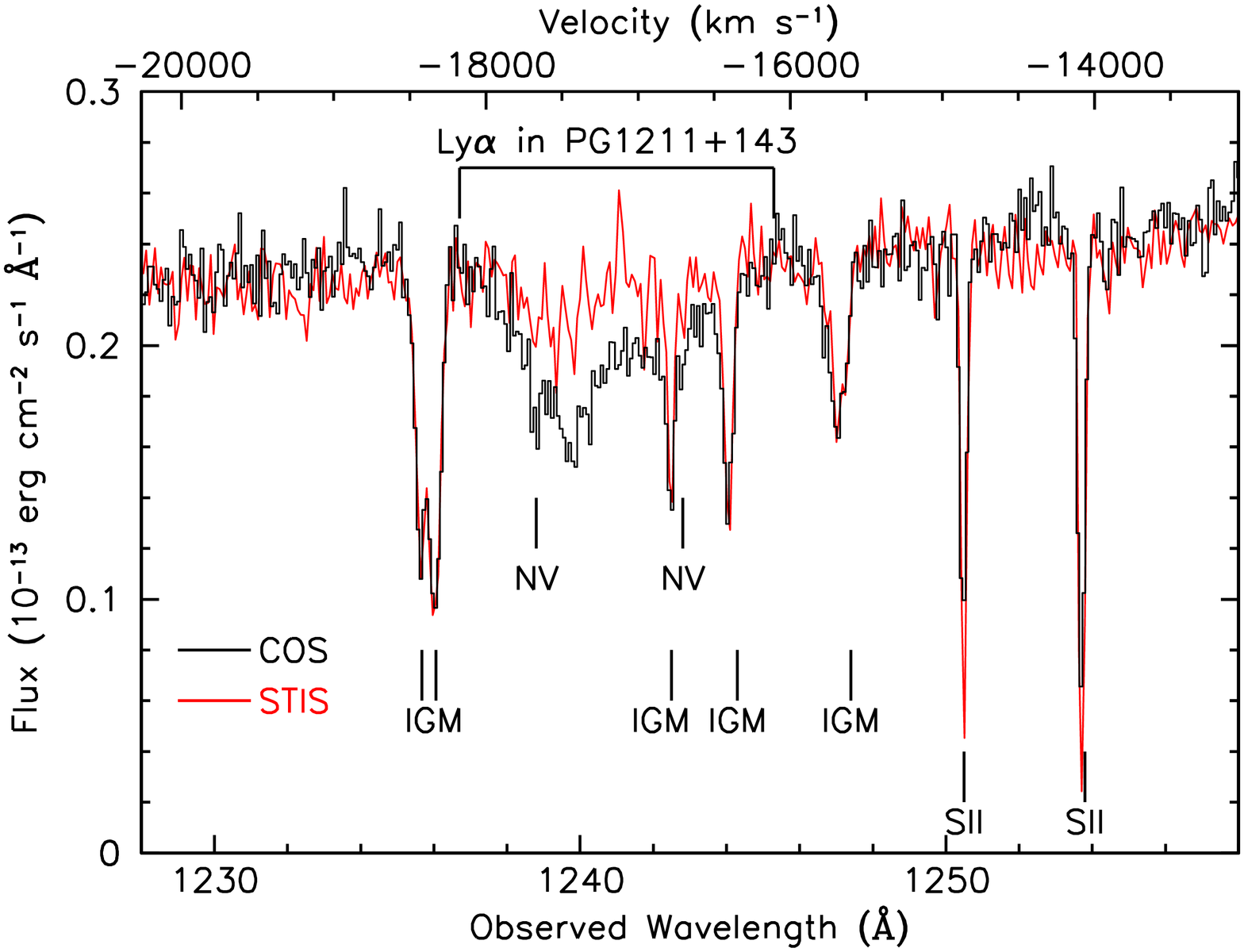}
\caption{Comparison of the STIS E140M spectrum of PG1211+143 to the
{\it HST}/COS G130M spectrum in the region around the broad Ly$\alpha$
absorption feature shows that the absorption feature has varied.
The red line is the STIS data binned by 6 pixels; the black histogram is the
COS spectrum binned by 8 pixels.
The lower horizontal axis is the observed wavelength in \AA ngstroms.
The upper horizontal axis gives the outflow velocity of Ly$\alpha$ in the
rest frame of PG1211+143 at $z = 0.0809$.
Interstellar and intergalactic lines are labeled as in Figure \ref{FigLya}.
}
\label{FigLyaVary}
\end{figurehere}
\vskip 12pt

In contrast to the outflow at $v_{\rm out} = -16\,980$ \kmps\ 
($z_{\rm out} = -0.0551$), seen
with both {\it Chandra} and {\it HST}-COS, in neither observation
do we detect the ultra-high velocity gas at $-0.129 c$ previously noted by
\cite{Pounds16}.
\cite{Pounds16} cite multiple transitions of \ion{Fe}{25} and \ion{Fe}{26}
as evidence for this higher velocity gas.
It has a total equivalent hydrogen column density
of $\rm N_H = (3.7 \pm 2.9) \times 10^{23}~cm^{-2}$ at an even higher
ionization parameter of $\rm log~\xi = 4.0$.
For the PG1211+143 spectral energy distribution in \cite{Danehkar17},
the fractional abundance of \ion{H}{1} scales with $\xi$ as
$\rm log~N_{HI} = -3.47 - 1.4~log ~\xi$.
Thus this $-0.129c$ gas component should have an associated neutral hydrogen
column density of $\rm\sim 3.1 \times 10^{14}~cm^{-2}$, which should be easily
visible in a UV spectrum. Indeed, this is as strong as the \lya\ 
absorption we detect that is associated with the lower velocity, lower
ionization component of the outflow detected in our {\it Chandra} spectrum.
At the location of a putative $-0.129 c$ component in our COS spectrum, we can
set an upper limit on any \lya\ absorption of
$< 1.0 \times 10^{14}~\rm cm^{-2}$,
well below any expected absorption associated with such a component.
At $\rm log~\xi = 3.4~or~4.0$,
the trace columns of other UV-absorbing ions such as
\ion{O}{6}, \ion{N}{5}, or \ion{C}{4} would be $\rm N_{ion} < 10^{13}~cm^{-2}$
and undetectable in our COS spectra.
Although we detect neither X-ray nor UV absorption associated with the
high-velocity $-0.129 c$ outflow, this could simply be due to variability, as
even the \ion{H}{1} counterpart of the
 $v_{\rm out} = -17\,300$ \kmps\ ($z_{\rm out} = -0.0561$) X-ray absorber
is not always detectable, as shown in Fig. \ref{FigLyaVary}.

While we do not confirm all of the ultrafast outflow components previously
seen in PG1211+143, we do have a robust detection of one at an outflow
velocity of $v_{\rm out} = -16\,980~\kmps$ ($z_{\rm out} = -0.0551$).
However, this outflow component is considerably
lower in total column density than previously suggested features.
Is it then massive enough and energetic enough to have a substantive impact
on the evolution of its host galaxy?
As usual, this is still ambiguous since derivation of the mass outflow rate
and the kinetic luminosity depend on the location of the absorbing gas we
have detected. The further from the central source, the more massive the
outflow, and the higher its kinetic luminosity.
Assuming the outflow is in the form of a partial thin spherical shell moving
with velocity v, its mass flux, $\dot M$, and kinetic luminosity, $\dot {E_k}$,
are given by:
{\center
\vskip -20pt
$$\rm \dot M = 4\pi \Delta\Omega R N_H \mu m_p {\rm v} $$
$$\rm \dot E_k = \frac{1}{2} \dot M {\rm v}^2 $$
}
\noindent
where $\Delta\Omega$ is the fraction of the total solid angle occupied by the
outflow, $\rm R$ is the distance of the outflow from the central source,
$\rm N_H$ is the total hydrogen column density of the outflow,
$\rm m_p$ is the mass of the proton, and
$\mu = 1.15$ is the molecular weight of the plasma per proton.
Since \cite{Tombesi10} argues that roughly 50\% of AGN have ultrafast,
high-ionization outflows, we assume $\Delta\Omega = 0.5$.

As many authors have argued, the maximum radius can be estimated by assuming
a plasma of uniform density distributed along the line of sight to the central
source, so that $\rm N_H = n R$ (e.g., \cite{Blustin05, Reeves12, Ebrero13}).
Given that we know the ionization parameter $\rm \xi = L_{ion} / (n R^2)$,
this gives the constraint $\rm R < L_{ion} / (N_H \xi)$.
For our observation of PG1211+143 and the SED presented by \cite{Danehkar17},
$\rm L_{ion} = 1.587 \times 10^{45}~\rm erg~s^{-1}$.
$\rm N_H = 3 \times 10^{21}~cm^{-2}$ and $\rm v = 16\,980~\rm \kmps$, so that
$\rm R < 265$ pc,
$\rm \dot M < 799~\Msun~yr^{-1}$ and
$\rm \dot E_k < 7.3 \times 10^{46}~\rm erg~s^{-1}$.
Our SED for PG1211+143 gives a bolometric luminosity of
$5.3 \times 10^{45}~\rm erg~s^{-1}$, so at this maximum distance the outflow
would be depositing up to $14 \times$ the bolometric luminosity as mechanical
energy into the host galaxy.
This even exceeds the Eddington luminosity
of $1.8 \times 10^{46}~\rm erg~s^{-1}$
for its black hole mass of $\rm 1.46 \times 10^8~\Msun$ \citep{Peterson04}.
However, given the unrealistic assumption involved in this approximation
(i.e., a single ionization parameter describes gas uniformly distributed from
0 to 265 pc), this merely demonstrates the potentially powerful influence of
this outflow on the host galaxy.

At the other extreme, if we assume the gas is a thin spherical shell at the
radius where its velocity equals the escape velocity of its central
black hole, for $\rm v = 16\,980~\rm \kmps$, $\rm R = 5$ lt-days,
the impact is minimal, with a mass outflow rate of $>0.013~\Msun~yr^{-1}$,
and a kinetic luminosity of $>1.2 \times 10^{42}~\rm erg~s^{-1}$.

If the absorbing cloud is associated with an ejection event in 2001, we can
set a better-motivated constraint on the location of the absorber.
Variability in the Ly$\alpha$ absorption feature argues for changes related
to motion of the absorber rather than an ionization response due to the
magnitude of the variability.
The observed changes in strength (from absence, or near-absence) from the prior
FOS and STIS observations to our COS observation are much stronger than
expected based on changes in the luminosity of PG1211+143.
In the archival record, PG1211+143 spans a range in
UV brightness at 1464 \AA\ from
$2.0 \times 10^{-14}~\rm erg~cm^{-2}~s^{-1}~\AA^{-1}$ to
$4.6 \times 10^{-14}~\rm erg~cm^{-2}~s^{-1}~\AA^{-1}$ \citep{Dunn06}.
The FOS and STIS observations bracket this range with flux levels of
$2.1 \times 10^{-14}~\rm erg~cm^{-2}~s^{-1}~\AA^{-1}$ and
$4.6 \times 10^{-14}~\rm erg~cm^{-2}~s^{-1}~\AA^{-1}$, respectively.
Our COS observation lies near the lower end at
$2.4 \times 10^{-14}~\rm erg~cm^{-2}~s^{-1}~\AA^{-1}$.
For our adopted SED \citep[Figure 4 in][]{Danehkar17}),
a factor of 2 change in flux
translates to a factor of 2.5 change in the neutral hydrogen column density.
Given the saturation present in Ly$\alpha$ in our COS spectrum, the
column density has increased by more than a factor of 3.5.
This bolsters the case that the X-ray/UV outflow is not continuous.
It could either have originated as an ejection event around 2001, or it
could imply that the absorbing cloud is moving transverse to
our line of sight. 

For motion at $-16\,980~\rm km~s^{-1}$, an ejected cloud would have moved
outward to a distance of $7 \times 10^{17}$ cm (0.23 pc) over 13 years.
A thin spherical shell at this distance would imply a mass outflow rate of
$>0.013~\Msun~yr^{-1}$,
and a kinetic luminosity of $6 \times 10^{43}~\rm erg~s^{-1}$.
It is interesting to observe that this is more similar to the minimum kinetic
luminosity of $>3 \times 10^{44}~\rm erg~s^{-1}$ we derived
for a potential radio jet based on the VLA observations \citep{Danehkar17},
which appears energetically similar to the X-ray/UV outflow.

Unfortunately, none of these estimates is definitive since we have no good
measurement of the actual location and duration of the outflow.
With so few spectral diagnostics, pinning down the radius of such an
outflow would require intensive monitoring to measure recombination timescales
in the photoionized gas.

\section{Summary}

We have obtained high S/N UV spectra of the QSO PG1211+143 covering the
900--1800 \AA\ bandpass simultaneously with a deep {\it Chandra} X-ray
observation \citep{Danehkar17}.
Our ultraviolet spectra detect a fast, broad \lya\ absorption
feature outflowing at a velocity of $v_{\rm out} = -16\,980~\kmps$
($z_{\rm out} = -0.0551$) with a FWHM of 1080 \kmps.
A possible feature associated with \lyb\ is also detected at 99.8\% confidence,
but no other ionic species are detected in absorption at this velocity.
This H\,{\sc i} absorption feature is a likely counterpart of the highly
ionized warm absorber detected in our {\it Chandra} HETGS spectrum at an
outflow velocity of $v_{\rm out} = -17\,300~\kmps$ ($z_{\rm out} = -0.0561$).
This ultrafast outflow may be the same as the $v_{\rm out} \sim -0.06 c$
outflow
reported in previous {\it XMM-Newton} observations by \cite{Pounds16} and
\cite{ Reeves17}.
Our detection of H\,{\sc i} absorption associated with these
outflows demonstrates that neutral hydrogen is a very sensitive tracer of
high-column density gas even at high ionization.

\acknowledgments
Based on observations made with
the NASA/ESA {\it HST}, and obtained from the Hubble
Legacy Archive. This work was supported by NASA
through a grant for {\it HST} program number 13947
from the Space Telescope Science Institute (STScI), which is
operated by the Association of Universities for Research
in Astronomy, Incorporated, under NASA contract
NAS5-26555, and from the Chandra X-ray Observatory (CXC) via grant GO5-16108X.
TF was partly supported by grant 11525312 from the
National Science Foundation of China.

\bibliographystyle{apj}
\bibliography{pg1211}

\end{document}